\documentclass[11pt]{article}
\usepackage{graphics,epsfig,psfrag,amsfonts,amssymb}
\begin{document}

  \makeatletter
  \@addtoreset{equation}{section}
  \makeatother
  \renewcommand{\theequation}{\thesection.\arabic{equation}}
  \baselineskip 15pt

\title{\bf On deciding whether a Boolean function is constant or not
\footnote{Work supported in part by Istituto Nazionale di Fisica Nucleare,
 Sezione di Trieste, Italy}}
\author{Fabio Benatti \footnote{e-mail: benatti@ts.infn.it}\\
 {\small Department of Theoretical Physics of the University of Trieste and}\\
 {\small Istituto Nazionale di Fisica Nucleare, Sezione di Trieste, Italy}\\
 \\ Luca Marinatto\footnote{e-mail: marinatto@ts.infn.it}\\
 {\small International Centre for Theoretical Physics ``Abdus Salam", Trieste
 and}\\
 {\small Istituto Nazionale di Fisica Nucleare, Sezione di Trieste, Italy}\\}

 \date{}

 \maketitle

\begin{abstract}
We study the probability of making an error if, by querying an
oracle
 a fixed number of times, we declare constant a randomly chosen $n$-bit
 Boolean function.
We compare the classical and the quantum case, and we determine for how many
 oracle-queries $k$ and for how many bits $n$ one querying procedure is more
 efficient than the other.
\end{abstract}



\section{Introduction}

\noindent \noindent Query complexity theory is mainly concerned
with the computational cost required to determine some specific
property of functions. The cost is measured by the number of
queries which can be addressed to a
 ``black box" device (the oracle) that outputs instantaneously an answer
to a query. Various results concerning query complexity of
Boolean functions,
 i.e. functions with a finite-size domain which have two possible output
 values only, exist in the literature~\cite{bbcmw,Gold,vanDam}.
The first of this sort of problems which has been addressed by the quantum
 computational theory, is represented by the Deutsch-Josza
 algorithm~\cite{dj}.
It involves the decision whether a given Boolean function of $n$
binary digits  is {\em constant} (that is whether it outputs the
same binary digit for every
 input value) or {\em balanced} (that is whether it outputs $1$ on exactly
 half
 inputs and $0$ on the other half) and it can be solved with certainty by
 querying a quantum oracle just once.

This task is successfully achieved by exploiting two essential features of
 quantum mechanics: its linearity, which allows the
 simultaneous evaluation of a function on linear superpositions of its input
 values, and the interference between amplitudes, which raises the
 probability
 of obtaining the desired results.

The {\em a-priori} knowledge that the function computed by the quantum
oracle
 is either constant or balanced, is crucial in solving the
 problem exactly, i.e. with zero probability of error.
In this paper, we address the question of deciding the constancy of a
generic
 function chosen at random among the set of $2^{2^{n}}$ $n$-bit Boolean
 functions. In such a case, the iteration of the same algorithmic procedure
 permits one
 to decide about the constancy of such generic function with a probability of
 error
 which depends on $n$ and on the number $k$ of quantum oracle-queries.

Unfortunately, this probability  tends to $1$ in the worst case
scenario, namely
 when the given function outputs the same value for every input but one.
Nevertheless, it is our purpose to determine the efficiency of
the quantum algorithm and the range of values of $n$ and $k$ for
which it is more efficient compared to the unique classical
procedure which involves querying the oracle successively until
two different outputs are encountered.

Not too surprisingly, due to classical conditioning of each output on the
 previous ones, in some specific situations (the worst cases), for
 large number of inputs and nearly as many oracle-queries, the efficiency
 of the classical algorithm turns out to overcome that of the quantum
 one. Moreover, in the average case, i.e. when the $n$-bit Boolean function
 is sorted at random, the quantum algorithm is always preferable to the
 classical one.


\section{The worst case analysis}

The general problem we are going to tackle is as follows. Suppose
that we are given an oracle which instantaneously
 computes an arbitrary n-bit Boolean function $f\,:\,\left\{
 0,1\right\}^{n} \rightarrow \left\{ 0,1\right\}$ on inputs that,
 as usual, will be enumerated by the integers
 $\left\{ 0,1,\dots, 2^{n}-1\right\}$.
Our goal is to devise a quantum algorithm, involving oracle
 queries, in order to determine, with some probability of error, whether
 such
 a function is constant or not.

Since, by hypothesis, we have no {\em a-priori} information or knowledge
 about $f$, we can only suppose that it has been sorted completely at random
 within the set ${\cal{S}}_{n}$ of $2^{2^{n}}$ n-bit Boolean functions.

If we have at our disposal a classical oracle which outputs the value of
$f$,
 and which can be queried at most $k$ times, the only conceivable classical
 algorithm able to solve the problem is$\:$: ``{\em query the oracle $k$
 times
 or until two output values are different: in the latter case stop
 querying and declare that the function is not constant, otherwise say that
 $f$ is constant}".

Of course, this simple algorithm fails to give the right answer only when
the
 function is actually non-constant and when we have obtained $k$
 consecutive equal digits in the querying procedure.
In fact, the function $f$ might output a different digit on any of the
untried
 inputs, so that one has to evaluate the probability of such an event.

We start by considering the worst possible scenario, that is the
unknown  function $f$, drawn from the set ${\cal{S}}_{n}$,
actually outputs the same
 bit, say $0$, for all of its inputs but one (for simplicity of notation we
 will say that such a function is of $f_{1}$-type).
The probability $p_{1}(k,n)$ of declaring erroneously constant
 such a function, coincides with the probability of obtaining $k$
 consecutive
 digits $0$ after querying the oracle $k$ times:
\begin{equation}
\label{worst1}
p_{1}(k,n)= \textrm{Prob} [\underbrace{0\dots 0}_{k}] = \textrm{Prob}[0]
\cdot
\Big( \prod_{\ell=1}^{k-1}\textrm{Prob}[\,0 |
\underbrace{0\dots 0\,}_{\ell}] \Big),
\end{equation}
where we have indicated with
$\textrm{Prob}[\,0|\underbrace{0\dots 0\,}_{\ell}]$, $1\leq \ell \leq k-1$,
 the conditional probability of getting $0$ after the $(\ell +1)$-th oracle
 query if a sequence of $\ell$ consecutive zeroes has already been obtained.
Since the function $f$ is of $f_{1}$-type, one easily calculates
$\textrm{Prob}[0]=(2^{n}-1)/2^{n}$
 and $\textrm{Prob}[0|\underbrace{0\dots0\,}_{\ell}]=
 (2^{n}-\ell-1)/(2^{n}-\ell)$, whence
\begin{equation}
\label{worst2}
p_{1}(k,n)= \prod_{j= 1}^{k} \frac{2^{n}-j}{2^{n}-(j-1)}=
1-\frac{k}{2^{n}}\:.
\end{equation}
This probability distribution is equal to zero for $k=2^{n}$; indeed, due to
 statistical correlations between subsequent outputs,
 in such a case we can determine with certainty if the function is constant
 or not.
As we will see, this is not the case with the Deutsch-Josza
quantum algorithm which, unless the function is either constant or
balanced, is not
 able to give the correct answer with certainty.
First, let us briefly recall the basics of that algorithm whose associated
 quantum circuit is the following:

\medskip
\begin{picture}(500,70)
 \put(120,40){\framebox(30,20)}
 \put(220,40){\framebox(30,20)}
 \put(120,0){\framebox(20,20)}
 \put(170,0){\framebox(30,60)}

 \put(110,10){\line(10,0){10}}
 \put(140,10){\line(20,0){30}} 
 \put(200,10){\line(20,0){60}}

 \put(110,50){\line(10,0){10}}
 \put(150,50){\line(60,0){20}} 
 \put(200,50){\line(10,0){20}}
 \put(250,50){\line(10,0){10}}

 \put(225,47){\large ${\bf H}^{\otimes\,n}$}
 \put(125,47){\large ${\bf H}^{\otimes\,n}$}
 \put(178,27){\large ${\bf U}_{f}$}
 \put(125,7){\large ${\bf H}$}

 \put(90,7){\large $\vert 1\rangle$}
 \put(80,47){\large $\vert 0\rangle^{\otimes n}$}
 \put(270,47){\large $\vert \psi\rangle$}
\end{picture}
\medskip

\noindent We have indicated with $H^{\otimes n}$ the $n$-fold Hadamard
 transformation on a $n$-qubit state and with $U_{f}$ the quantum oracle
 associated to $f$.
It is defined by
 $U_{f}\vert x\rangle \vert y \rangle = \vert x \rangle \vert y\oplus
 f(x)\rangle$, where $\oplus$ denotes the sum modulus $2$, $\vert x \rangle$
 is an $n$-qubit state such that $x\in \left\{ 0,1,\dots, 2^{n}-1\right\}$,
 and $\vert y \rangle$ is a one-qubit state.
If the initial input states are those indicated in the figure,
 it is possible to prove~\cite{dj} that the $n$-qubit output state
 $\vert \psi \rangle$ in the figure above, turns out to be$\,$:
\begin{equation}
\label{worst3}
\vert \psi \rangle = \sum_{z=0}^{2^{n}-1}\Big(
\sum_{x=0}^{2^{n}-1} \frac{(-1)^{z\cdot x +f(x)}}{2^{n}} \Big)
\,\vert z\rangle\:,
\end{equation}
where $z\cdot x$ is the bitwise inner product of $z$ and $x$, i.e.
 $z\cdot x= \sum_{j=1}^{n}z_{j}x_{j}$.
Therefore, according to Eq.~(\ref{worst3}),
 a measurement process performed onto the $n$-qubit computational
 basis $\left\{ \vert z\rangle \right\}$ yields the result
 $z=0$ with the following probability$\,$:
\begin{equation}
\label{worst4}
\textrm{Prob}[z=0, \vert \psi \rangle] = \Big\vert \frac{1}{2^{n}}
\sum_{x=0}^{2^{n}-1} (-1)^{f(x)}\Big\vert^{2}\:.
\end{equation}
If the function $f$ is constant, the result $z=0$ is obtained with
certainty;
 this in turn implies that, if any other value $z\neq 0$ is obtained, the
 function cannot be constant.
Yet, if the function is of $f_{1}$-type, the result $z=0$ is
 obtained with a probability strictly smaller than $1$, given by:
\begin{equation}
\label{worst4.5}
\textrm{Prob}[z\!=\!0, \vert \psi \rangle \,{\bf \vert}\, f\!=\!f_{1}] =
\Big\vert \frac{1}{2^{n}}
\sum_{x=0}^{2^{n}-1} (-1)^{f(x)}\Big\vert^{2}=
 \Big\vert \frac{1}{2^{n}} (-2+2^{n})\Big\vert^{2}=
\Big( 1-\frac{1}{2^{n-1}} \Big)^{2}\:.
\end{equation}
The simplest quantum procedure to determine whether the function $f$
 is constant or not, consists of a simple iteration of the quantum circuit
 described above and it amounts to: ``{\em iterate the
 Deutsch-Jozsa algorithm $k$ times or until a measurement result $z\neq 0$
 appears: in the latter case stop querying and declare that the
 function is not constant, otherwise say that $f$ is constant}".

We denote by $q_{1}(k,n)$ the probability that, after obtaining $z=0$ in $k$
 consecutive measurements, we are wrong in asserting the constancy
 of the $f_{1}$-type function.
According to Eq.~(\ref{worst4.5}), one gets:
\begin{equation}
\label{worst5}
q_{1}(k,n)= \Big( 1-\frac{1}{2^{n-1}} \Big)^{2k}\:.
\end{equation}
It is worth noting that, contrary to the classical case where the
 probability of obtaining the same output value changes, by conditioning,
 after each oracle query, the consecutive (quantum) oracle-queries are
 independent from each other.
Therefore, the probability of obtaining $z=0$, $k$ times consecutively, is
 the product of the probabilities of the uncorrelated events; quantum
 conditioning on past results would certainly improve the efficiency of the
 algorithm, however we do not know of any such technique.

In order to compare the quantum querying procedure
 with respect to the classical one in this worst case, we define the
 following efficiency function $\Delta_{1}(k,n)$:
\begin{equation}
\label{worst6}
\Delta_{1}(k,n)\equiv p_{1}(k,n)- q_{1}(k,n)=
1-\frac{k}{2^{n}}- \Big( 1-\frac{1}{2^{n-1}} \Big)^{2k}\:.
\end{equation}
Given $n$ (the number of bits) and $k$ (the number of oracle-queries),
 a positive value of $\Delta_{1}(k,n)$ indicates that the quantum algorithm
 is more efficient than the classical one.
For $n \geq 2$ and $ 1\leq k \leq 2^{n}-1$, $\Delta_{1}(k,n)$ attains its
absolute
 maximum at $n=2$ and $k=1$:
\begin{equation}
 \label{worst7}
 \max_{k,n}\:{\Delta_{1}(k,n)} = \Delta_{1}(k=1,n=2) = 0.5\:\:.
\end{equation}
For fixed $n$, the function $\Delta_{1}(k,n)$ has relative maxima at
\begin{equation}
 \label{worst8}
 k^{\star} \cong 0.5\, \frac{ \log{\Big( -
 \frac{2^{-1-n}}{\log{(1-2^{1-n})}}}\Big) }{\log{(1-2^{1-n})}}\:
 \approx \:\:0.35 \cdot 2^{n}\:\:\:\:\:\: \textrm{for} \:\:2^{n}\gg 1\:,
\end{equation}
with $\Delta_{1}(k^{\star},n) \approx 0.40$, for $2^{n}\gg 1$.

In Fig.~\ref{fig1}, $\Delta_{1}(k,n)$ is plotted as a function of $k$ for
 different values of $n$; the curves displayed from left to right are
 associated with increasing values of $n$, $5 \leq n \leq 12$:

\begin{figure}[th]
\centerline{\epsfig{file=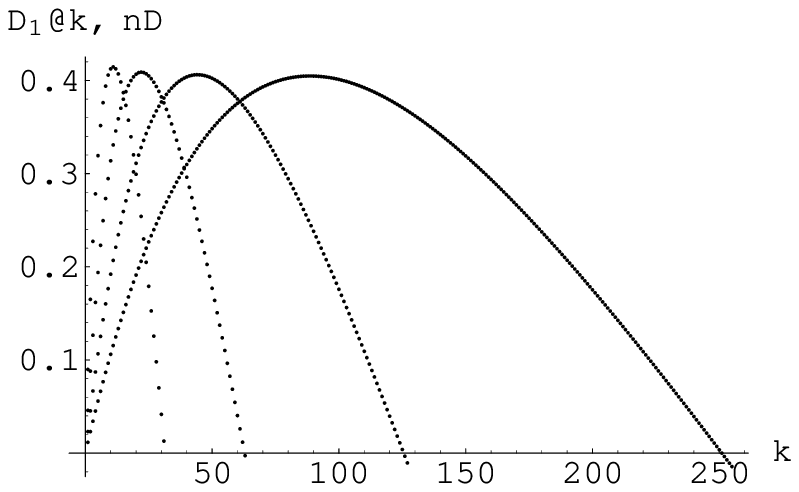, width=6cm} \hspace{40pt}
\epsfig{file=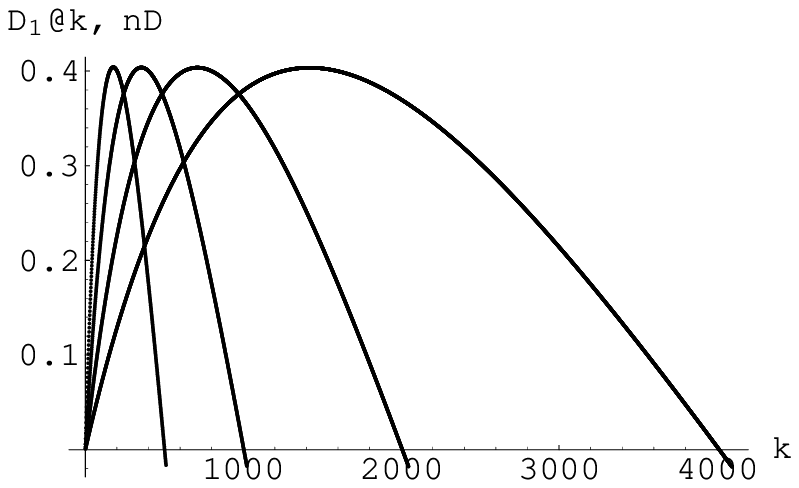, width=6cm}} \vspace*{8pt}
\caption{\label{fig1}Efficiency functions $\Delta_{1}(k,n)$
plotted
 for different values of n, $5\leq n \leq 8$
 on the left, $9\leq n \leq 12$ on the right and $1 \leq k \leq 2^{n}-1$.}
\end{figure}

As it can be argued from the picture plotted above,
 $\Delta_{1}(k,n)$ is a concave function in the variable $k$ for every fixed
 value of $n\geq 3$.
Therefore, the worst case efficiency of the quantum algorithm increases,
 with respect to the classical one, for $k \leq k^{\star}$ (where
 $k^{\star} \approx 0.35\cdot 2^{n}$ for $2^{n}\gg 1$), until it reaches
 its (relative) maximum value.
Such a value exhibits a weak dependence on the variable $n$ and it is
 approximately equal to $0.40$ for $2^{n}\gg 1$.
On the contrary, for $k > k^{\star}$, $\Delta_{1}(k,n)$ decreases to zero,
 thus indicating a substantial equivalence of the two methods for large
 $k$'s.

The above results show that, in the worst case, it is almost
always better
 to resort to the quantum rather than to the classical algorithm.
However, the improvement of quantum queries over the classical
ones decreases with the number of queries. Furthermore,
$\Delta_{1}(k,n)$ becomes negative for $k \cong 2^{n}$, namely
 there is a (small) range of values of $k$ for which the classical
 querying procedure gives less probability of error than the quantum one.

Not surprisingly, this is due to the fact that the $2^{n}$-th classical
 query is able to determine with certainty if the function is constant
 or not, since $p_{1}(k=2^{n},n)=0$.
Instead, the quantum algorithm, as it has been conceived, can never answer
 with certainty since $q_{1}(k,n)$ never vanishes.


\section{The average case analysis}

Until now we have restricted our analysis to the worst possible
scenario, i.e. when the function $f$ outputs the same value for
all possible inputs but one. In the general case, the function may
be  any of the $2^{2^{n}}$ possible $n$-bit Boolean functions
belonging to the set
 ${\cal{S}}_{n}$.
This set contains $2$ constant functions (one always outputs $0$,
 while the other outputs $1$), ${2^{n} \choose 2^{n-1}}$ balanced functions
 (they output the same value for exactly one-half of the possible inputs)
 and $2\cdot {2^{n} \choose 2^{n}-m}$ $f_{m}$-type functions (they output
 $0$ ($1$) on $2^{n}-m$ inputs and $1$ ($0$) on the remaining $m$ ones).

We repeat the analysis of the previous section and suppose we are
 given one particular $f_{m}$-type function.
The probability of obtaining always the same output bit by querying the
 classical oracle $k$ times, and thus of concluding, erroneously, that the
 function is constant, is equal to $\textrm{Prob} [\underbrace{0\dots
 0}_{k}]
+ \textrm{Prob} [\underbrace{1\dots 1}_{k}]$. The latter are the
probabilities of obtaining $k$ consecutive digits
 $0$ and $1$ respectively, and they explicitly read
 \begin{equation}
 \label{average1}
   \textrm{Prob} [\underbrace{0\dots 0}_{k}] =
  \prod_{j=0}^{k-1} \frac{2^{n}-m-j}{2^{n}-j}=
  \frac{(m-2^{n})_{k}}{(-2^{n})_{k}}\:,
 \end{equation}
 \begin{equation}
 \label{average2}
  \textrm{Prob} [\underbrace{1\dots 1}_{k}] =
  \prod_{j=0}^{k-1} \frac{m-j}{2^{n}-j}=
  \frac{(-m)_{k}}{(-2^{n})_{k}}\:.
 \end{equation}
In the above, for simplicity of notation, the Pochhammer symbol
 $(a)_{n} \equiv a(a+1)\dots (a+n-1)$ has been used.
Therefore, the classical error probability $p_{m}(k,n)$ in declaring
constant
 an $f_{m}$-type function is:
 \begin{equation}
 \label{average2.5}
  p_{m}(k,n) = \frac{(m-2^{n})_{k}}{(-2^{n})_{k}}
  +\frac{(-m)_{k}}{(-2^{n})_{k}}.
 \end{equation}
The corresponding probability $q_{m}(k,n)$ for the quantum case is much
easier
 to calculate.
In fact, given an $f_{m}$-type function, the conditional
 probability of obtaining $k$ consecutive measurement results $z=0$ via the
 iterated Deutsch-Jozsa procedure is, according to Eq.~(\ref{worst4}):
\begin{equation}
\label{average4}
 \textrm{Prob}[k \:\textrm{times} \:z\!=\!0,{\bf |}\,
  f\!=\!f_{m}]  \equiv q_{m}(k,n) = \Big\vert \frac{1}{2^{n}}
\sum_{x=0}^{2^{n}-1} (-1)^{f(x)}\Big\vert^{2k}=
\Big( 1-\frac{m}{2^{n-1}} \Big)^{2k}\:.
\end{equation}
As in the previous section, we can now define a function
 $\Delta_{m}(k,n)$ which quantifies the relative efficiency of the quantum
 algorithm with respect to the classical one:
\begin{equation}
\label{average4.5}
 \Delta_{m}(k,n) \equiv p_{m}(k,n) - q_{m}(k,n)
 =\frac{(m-2^{n})_{k}}{(-2^{n})_{k}} +\frac{(-m)_{k}}{(-2^{n})_{k}} -
\Big( 1-\frac{m}{2^{n-1}} \Big)^{2k}\: .
 \end{equation}
$\Delta_{m}(k,n)$ does not dramatically change with $n$; therefore, we fix
 $n\!=\!7$ and present typical behaviours for different values of $m$.

\begin{figure}[th]
\centerline{\epsfig{file=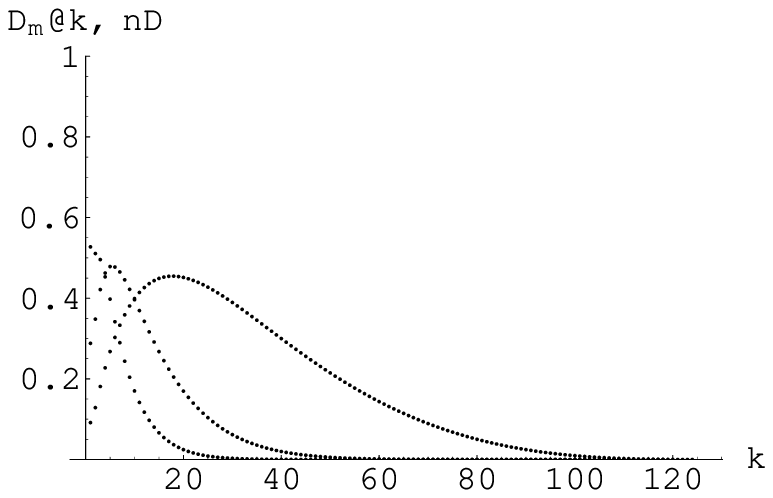, width=6cm} \hspace{40pt}
\epsfig{file=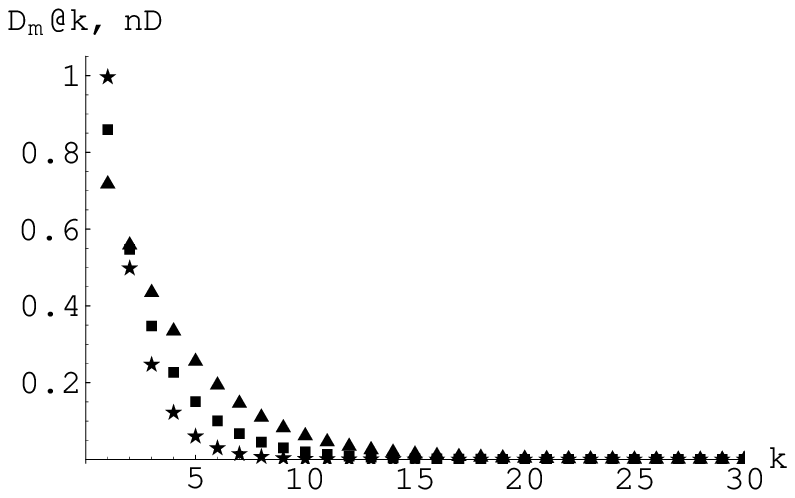, width=6cm} } \vspace*{8pt}
\caption{\label{fig2} Efficiency functions $\Delta_{m}(k,n)$ for
 $n=7$ and $m=3,10,20$ from right to left in the first
 picture and $m=30 (\blacktriangle),40 (\blacksquare), 60 (\bigstar)$
 in the second one.}
\end{figure}

We notice that $\Delta_{m}(k,n)$ is symmetric around $m=2^{n-1}$, and that,
 when $m$ gets close to $2^{n-1}$, the function $f_{m}$ tends to be balanced
 and $\Delta_{m}(k,n)$ rapidly decreases.
This behaviour indicates that the quantum algorithm can perform
much better than the classical one, but only for small numbers 
of oracle-queries. On the contrary, the two procedures become equally 
 efficient for larger values of $k$. For instance, in Fig.~\ref{fig2}, when
 $m=60$
and the function $f_{m}$ is almost balanced, the two procedures
are nearly indistinguishable for $k > 5$.
Instead, in the general case of $m$ significantly different from
 $2^{n-1}$, the dependence of $\Delta_{m}(k,n)$ on $k$
 is much smoother and presents a maximum and a change of concavity.
Further, for $k \cong 2^{n}$, $\Delta_{m}(k,n)$ assumes negative values,
 invisible in the above pictures.
This indicates a better performance of the classical algorithm
over the  quantum one, as already noticed when dealing with the
$f_{1}$-type function in the previous section. However this
phenomenon is much less pronounced than in the worst case
 situation.


\subsection{Randomly chosen Boolean functions}

We now tackle the general case and suppose we are given a
randomly chosen  Boolean function belonging to ${\cal{S}}_{n}$.
Then, the average classical error probability $\bar{p}(k,n)$ in
declaring such
 a function constant, after $k$ consecutive queries with identical outputs,
 is:
 \begin{equation}
 \label{average3}
 \bar{p}(k,n)=
 2\cdot\bigg[ \sum_{m=1}^{2^{n-1}-1}\, \frac{{2^{n}\choose m}}{2^{2^{n}}}
 \bigg(
  \frac{(m-2^{n})_{k}}{(-2^{n})_{k}}  + \frac{(-m)_{k}}{(-2^{n})_{k}}
  \bigg)\bigg] + \frac{{2^{n}\choose 2^{n-1}}}{2^{2^{n}}}\,
 \frac{2(-2^{n-1})_{k}}{(-2^{n})_{k}}\:.
 \end{equation}
In the previous expression, the factor ${2^{n}\choose m}/2^{2^{n}}$, with
 $m\neq 2^{n-1}$, gives the probability of sorting, within the set
 ${\cal{S}}_{n}$, an unbalanced $f_{m}$-type function which outputs a number
 of zeroes less than $2^{n-1}$ (the extra factor $2$ in front of the sum
 takes into account the contribution coming from functions which output a
 number of zeroes which is greater than $2^{n-1}$).
Analogously, the factor ${2^{n}\choose 2^{n-1}}/2^{2^{n}}$ gives the
 probability of sorting a balanced function.

On the contrary, the corresponding quantum error probability $\bar{q}(k,n)$
 of choosing at random a Boolean function within ${\cal{S}}_{n}$ and
 declaring
 it constant, due to $k$ consecutive measurement results $z=0$, is:
\begin{equation}
 \label{average5}
 \bar{q}(k,n)=
 \sum_{m=1}^{2^{n}-1}\, \frac{{2^{n}\choose m}}{2^{2^{n}}}
 \Big( 1-\frac{m}{2^{n-1}} \Big)^{2k}\:.
\end{equation}
As before, we introduce the average efficiency function $\bar{\Delta}(k,n)$:
\[
 \bar{\Delta}(k,n) \equiv \bar{p}(k,n)- \bar{q}(k,n)=  
 2\cdot\bigg[ \sum_{m=1}^{2^{n-1}-1}\, \frac{{2^{n}\choose m}}{2^{2^{n}}}
 \bigg(\frac{(m-2^{n})_{k}}{(-2^{n})_{k}} 
  + \frac{(-m)_{k}}{(-2^{n})_{k}}
  \bigg)\bigg] + \]
  \begin{equation}
 \label{average6}
 \frac{{2^{n}\choose 2^{n-1}}}{2^{2^{n}}}\,
 \frac{2(-2^{n-1})_{k}}{(-2^{n})_{k}} - 
   \sum_{m=1}^{2^{n}-1}\, \frac{{2^{n}\choose m}}{2^{2^{n}}}
 \Big( 1-\frac{m}{2^{n-1}} \Big)^{2k}\:.
\end{equation}
This function quantifies the (average) relative efficiency of the quantum
 algorithm with respect to the classical one: the larger
 $\bar{\Delta}(k,n)$,
 the more preferable the quantum procedure with respect to the classical
 one.

The dependence of $\bar{\Delta}(k,n)$ on $k$ and $n$ is shown in
Fig.~\ref{fig3} below, where its different behaviour with respect to
 $\Delta_{m}(k,n)$ in Fig.~\ref{fig2} emerges.

\begin{figure}[th]
\centerline{\epsfig{file=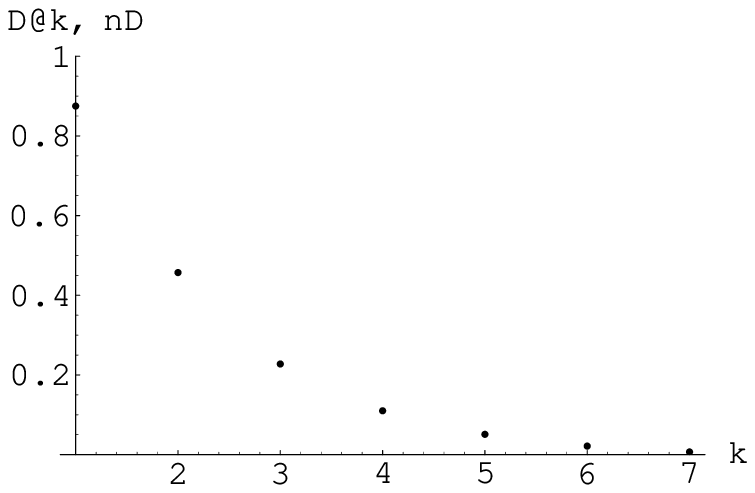, width=4.3cm} \hspace{15pt}
\epsfig{file=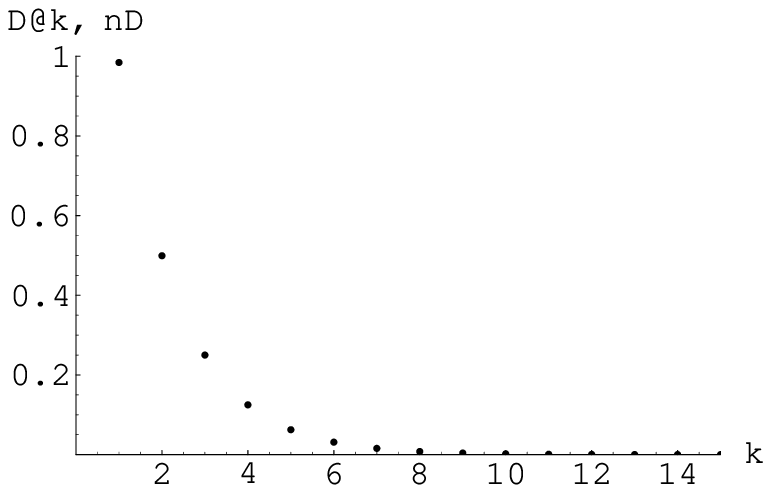, width=4.3cm} \hspace{15pt}
\epsfig{file=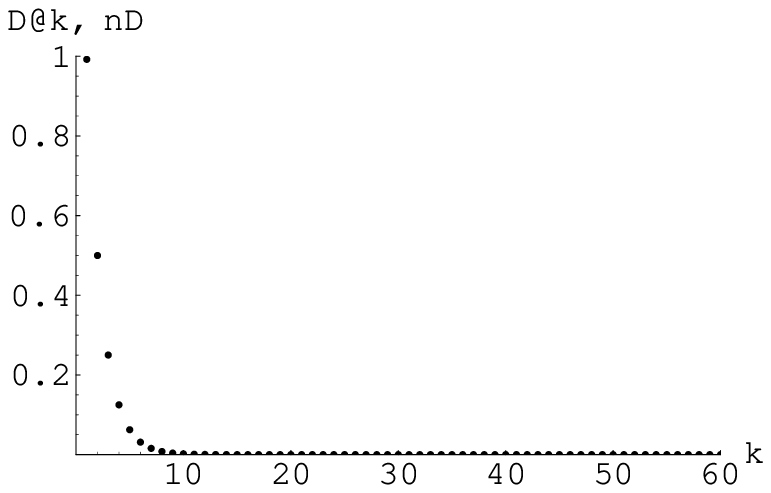, width=4.3cm} } \vspace*{8pt}
\caption{\label{fig3}Average efficiency functions
$\bar{\Delta}(k,n)$ for
 fixed values of $n=3,6,7$ from left to right, and
 $1 \leq k \leq 2^{n}-1$.}
\end{figure}

It is worth commenting on the behaviour of $\bar{\Delta}(k,n)$.
First, we notice that this function is approximately zero for most of the
 values of $n$ and $k$, indicating a substantial equivalence of the two
 algorithms; indeed, in all but a few situations, they are able to solve the
 problem with negligible probability of error.

Second, for fixed $n$, $\bar{\Delta}(k,n)$ is monotonically decreasing in
$k$;
for $k=1$, the quantum error probability $\bar{q}(k=1,n)$ falls rapidly
 to zero, while the classical one, $\bar{p}(k=1,n)$, approaches unity.

The decrease to zero of $\bar{q}(k=1,n)$ is due to the fact that the
 number of balanced (or almost balanced) functions increases considerably
 for
 $n$ large and the quantum algorithm is able, as we already know, to
 recognize
 them with certainty (or almost certainty).
On the contrary, the number of {\em worst} case functions, i.e.
those which
 output almost always the same value, gets rapidly to zero and the
 probability
 of sorting one of them becomes negligible.
This argument does not apply to the classical probability $\bar{p}(k=1,n)$,
 which instead tends to $1$; indeed, the classical algorithm fails to give
 the right answer for balanced (or almost balanced) functions which form
 the overwhelming majority in ${\cal S}_{n}$.

Finally, unlike the previous cases, the numerical analysis
indicates  no negative values of $\bar{\Delta}(k,n)$; thus, in
the average case,
 the quantum algorithm is expected to be always better than the classical
 one.


\section{Conclusions}

In this paper we have shown how a trivial iteration of the original
 Deutsch-Jozsa algorithm, can be applied to solve the general problem of
 deciding if a function $f$, sorted at random among the set of $n$-bit
 Boolean functions, is constant or not.
We have compared the error probability of the quantum procedure with that
 of the classical one, the latter consisting of consecutive oracle-queries
 performed until two different outputs are encountered.
We have analyzed the range of values of $n$ and $k$ for which the
 quantum procedure is more efficient than the classical one, both on
 average and in the worst possible situation.
From a numerical analysis, one concludes that, by iterating the quantum
 algorithm, one can always solve the problem, on average, {\em better}
 than with the classical method,
 and {\em much better} in those situations where the number of queries $k$
 is small and $2^{n}$ large.
Moreover, we have seen that, when $f$ is a $f_{m}$-type function
 with $m$ not much greater than $1$ (the worst cases), then
 the classical method is asymptotically (that is, when $2^{n}$ is large and
 $k\cong 2^{n}$) preferable to the quantum one.
This peculiar effect is due to the fact that the Deutsch-Jozsa
algorithm,
 unlike the classical querying, can never solve the general problem with
 certainty since it does not allow for conditioning on previous results.

%
%


\end{document}